\pdfoutput=1

\documentclass[11pt,twoside,a4paper,cmspaper,final,collab]{cms-tdr}
\def\svnVersion{179509}\def\cmsCernNo{2013-002}\def\cmsCernDate{\today}\def\cmsMessage{Submitted to Physical Review D}

\begin{document}\cmsNoteHeader{EXO-11-010}

\hyphenation{had-ron-i-za-tion}
\hyphenation{cal-or-i-me-ter}
\hyphenation{de-vices}
\RCS$Revision: 161251 $
\RCS$HeadURL: svn+ssh://svn.cern.ch/reps/tdr2/papers/EXO-11-010/trunk/EXO-11-010.tex $
\RCS$Id: EXO-11-010.tex 161251 2012-12-12 23:39:26Z prosper $

\ifthenelse{\boolean{cms@external}}{\providecommand{\cmsLeft}{top}}{\providecommand{\cmsLeft}{left}}
\ifthenelse{\boolean{cms@external}}{\providecommand{\cmsRight}{bottom}}{\providecommand{\cmsRight}{right}}
\cmsNoteHeader{EXO-11-010}

\title{\texorpdfstring{Search for contact interactions using the inclusive jet \pt spectrum in {\Pp\Pp} collisions at $\sqrt{s}=7$\TeV}{Search for contact interactions using the inclusive jet pt spectrum in pp collisions at sqrt(s) = 7 TeV}}

\date{\today}

\abstract{Results are reported of a search for a deviation in the jet
production cross section from the prediction of
perturbative quantum chromodynamics at next-to-leading order.
The search is conducted using a 7\TeV proton-proton data sample
corresponding to an integrated luminosity of 5.0\fbinv,
collected with the Compact Muon Solenoid detector
at the Large Hadron Collider.
A deviation could arise from interactions characterized by a mass scale
$\Lambda$ too high to be probed directly at the LHC.  Such phenomena can be
modeled as contact interactions. No evidence of a deviation is found.
Using the $\text{CL}_\text{s}$ criterion,
lower limits are set on $\Lambda$ of 9.9\TeV and 14.3\TeV at 95\% confidence
level for models with destructive and constructive interference,
respectively. Limits obtained with a Bayesian method are also reported.}

\hypersetup{%
pdfauthor={CMS Collaboration},%
pdftitle={Search for contact interactions using the inclusive jet pT spectrum in pp collisions at sqrt(s)=7 TeV},%
pdfsubject={CMS},%
pdfkeywords={CMS, physics, contact interactions, jets}}

\maketitle 

\section{Introduction}
\label{sec:intro}
Interactions at an energy scale much lower than the mass of the mediating particle
can be modeled by contact interactions (CI)~\cite{PhysRevLett.50.811,
RevModPhys.56.579, Chiappetta1991489, Lane:1996gr} governed by a
single mass scale conventionally denoted by $\Lambda$. A search for
contact interactions is therefore a search for interactions whose
detailed characteristics become manifest only at higher
energies.  Contact interactions
can affect the jet angular distributions as well as the jet transverse
momentum ($\pt$) spectra, particularly for low-rapidity jets.
Lower limits on $\Lambda$ have been set by the
CDF~\cite{PhysRevLett.79.2198}, \DZERO~\cite{PhysRevLett.82.4769}, and
ATLAS~\cite{PhysRevD.87.015010} collaborations.
The Compact Muon Solenoid (CMS) collaboration has previously
measured the dijet angular distribution~\cite{Chatrchyan:2012bf}
using a data set of $\sqrt{s}=$ 7\TeV proton-proton
collisions corresponding to an integrated luminosity of 2.2\fbinv, and found
$\Lambda > 8.4$\TeV and $\Lambda > 11.7$\TeV at 95\% confidence level (CL),
for models with destructively and constructively interfering amplitudes, respectively.

The inclusive jet $\pt$ spectrum, i.e., the spectrum
of jets in $\Pp + \Pp
\to \text{jet} + X$ events, where $X$ can be
any collection of
particles, is generally considered to be less sensitive to the presence
of contact interactions than the jet angular distribution.
This perception
is due to the jet $\pt$ spectrum's
greater dependence on the jet energy scale (JES) and on the parton
distribution functions (PDF), which are difficult to determine accurately.
However, considerable
progress has been made by the CMS collaboration in understanding the
JES~\cite{JME-10-011-PAS}.  The understanding of
PDFs has also improved
greatly at high parton momentum fraction~\cite{Nadolsky:2008zw,Ball:2011mu,Watt:2012tq},
in part because of the important constraints on the gluon PDF
provided by measurements at the Tevatron~\cite{Abazov:2011vi, Aaltonen:2008eq}.
These developments have made the jet $\pt$ spectrum
a competitive observable to search for phenomena described by
contact interactions, reprising
the method that was used in searches by CDF~\cite{Abe:1996wy}
and \DZERO~\cite{PhysRevD.62.031101}.

In this paper, we report the results of a search for a deviation in the
jet production cross section from the next-to-leading-order (NLO) quantum
chromodynamics (QCD) prediction of  jets produced at low-rapidity
with transverse momenta $>$500\GeV. The analysis is based on
a 7\TeV proton-proton data sample corresponding to
an integrated luminosity of 5.0\fbinv, collected with the CMS
detector at the Large Hadron Collider (LHC).

\section{Theoretical models}
\label{sec:ci} The experimental results are interpreted in terms of a
CI model described by the effective
Lagrangian~\cite{Chiappetta1991489, PhysRevD.67.115011}
\begin{equation} L = \zeta \, \frac{2\pi}{\Lambda^2} (\cPaq_L
\gamma^{\mu} \cPq_L)(\cPaq_L \gamma_{\mu} \cPq_L),
\label{eq:CImodel}
\end{equation} where $\cPq_L$ denotes a left-handed quark field and
$\zeta = +1$ or $-1$ for destructively or constructively interfering
amplitudes, respectively.  The amplitude for jet production can be
written as $$a = a_\text{SM} + \lambda \, a_\text{CI} \,$$
where $a_\text{SM}$ and $a_\text{CI}$ are the standard model (SM) and
contact interaction amplitudes, respectively.  Since the
amplitude is linear in $\lambda = 1/\Lambda^2$, the cross
section $\sigma_k$ in the
$k$th jet $\pt$ bin is given by
\begin{equation} \sigma_k = c_k + b_k \, \lambda + a_k \, \lambda^2,
\label{eq:sigmak}
\end{equation} where $c_k$, $b_k$, and $a_k$ are jet-$\pt$-dependent
coefficients.

We use models characterized
by the cross section $\text{QCD}_\text{NLO} + \text{CI}(\Lambda)$,
where $\text{QCD}_\text{NLO} = c_k$ is the inclusive jet cross section computed
at next-to-leading order, and $\text{CI}(\Lambda) = b_k \, \lambda + a_k \, \lambda^2$ parameterizes the
deviation of the inclusive jet cross
section from the QCD prediction arising from the hypothesized contact interactions.
The $\text{QCD}_\text{NLO}$ cross section is calculated with version 2.1.0-1062 of the
fastNLO program with scenario table fnl2332y0.tab~\cite{Kluge:2006xs}
using the NLO CTEQ6.6 PDFs~\cite{CTEQ}. We
do not unfold the observed inclusive jet $\pt$
spectrum.  Instead, the NLO QCD jet $\pt$ spectrum is
convolved with the CMS
jet response function, where the jet energy
resolution (JER) $\sigma_{\pt}$ for low-rapidity jets is given by
\begin{eqnarray}
\sigma_{\pt} & = & \pt \sqrt{ -\frac{n^2}{\pt^2} + \frac{s^2 \pt^m}{\pt} + c^2},
\label{eq:jer}
\end{eqnarray}
with  $n = 5.09$, $s = 0.512$, $m = 0.325$, $c = 0.033$,
and compared directly with the observed spectrum using a likelihood function.
Equation~(\ref{eq:jer}) is the standard form for the calorimeter resolution function, modified to account for a weak $p_\text{T}$ dependence of the coefficient of the ($p_\text{T}^{-1}$) stochastic term and to model better the  resolution of  low $p_\text{T}$ jets by using a negative coefficient
for the ($p_\text{T}^{-2}$) noise term.
For brevity, we shall refer to the smeared spectrum as the NLO QCD jet $\pt$ spectrum.

The signal term $\text{CI}(\Lambda)$ is modeled by
subtracting the leading-order (LO) QCD jet cross section $(\text{QCD}_\text{LO})$ from the
LO jet cross section computed with a contact term.
The leading-order jet $\pt$ spectra are computed by generating
events with and without a
CI term using the program \PYTHIA 6.422, the Z2 underlying event
tune~\cite{PhysRevD.67.115011, Sjostrand:2006za},
and the same CTEQ PDFs used to calculate $\text{QCD}_\text{NLO}$.
The generated events are
processed with the full CMS detector simulation program,
based on \GEANTfour~\cite{Agostinelli:2002hh}.  Interactions
between all quarks are included (Appendix~\ref{sec:CIgen}) and we consider
models both with destructive and
constructive interference between the QCD and CI amplitudes.
We note that NLO corrections to the contact interaction model have recently
become available~\cite{Gao:2011ha}, and we plan to use these results in future studies.
These corrections are expected to change the results by less than 5\%.

\begin{figure}[htb!]
\begin{center}
\includegraphics[width=0.49\textwidth]{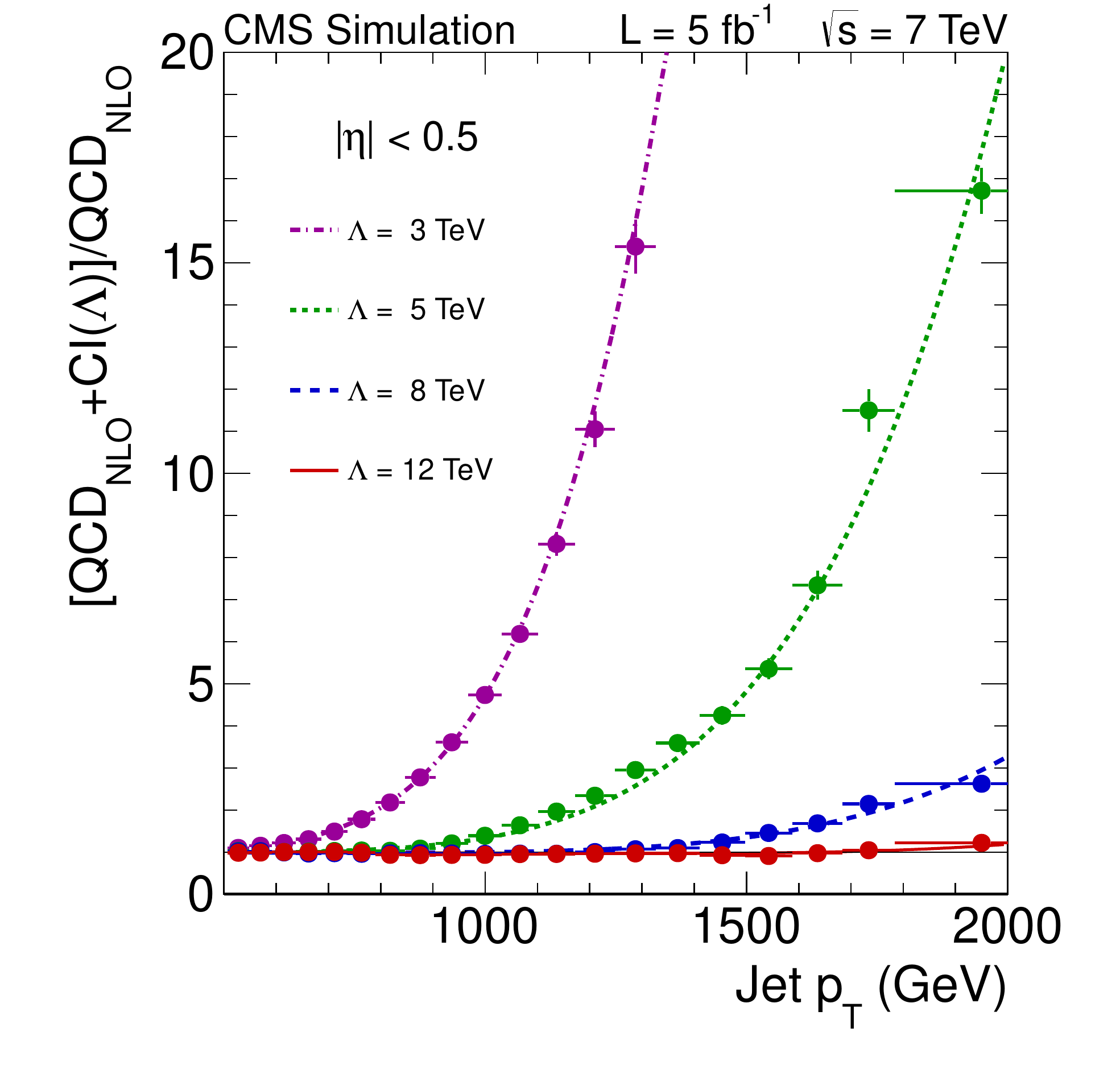}
\caption{
The cross section ratios, $f = [\text{QCD}_\text{NLO} + \text{CI}(\Lambda)]/\text{QCD}_\text{NLO}$, with $\Lambda = 3, 5, 8,$ and 12\TeV.
The points with error bars are the theoretical values of the cross section ratios.
The curves are the results of a fit of Eq.~(\ref{eq:f}) simultaneously to the four cross section ratios. The
NLO QCD jet $\pt$ spectrum is calculated
using the nominal values of the JES, JER,
PDF, renormalization and factorization scales for models
with destructive interference. The values of the parameters of the fit are given in Table~\ref{tab:fit}.
}
\label{fig:fit}
\end{center}
\end{figure}

\begin{figure}[htb!]
\begin{center}
\includegraphics[width=0.49\textwidth]{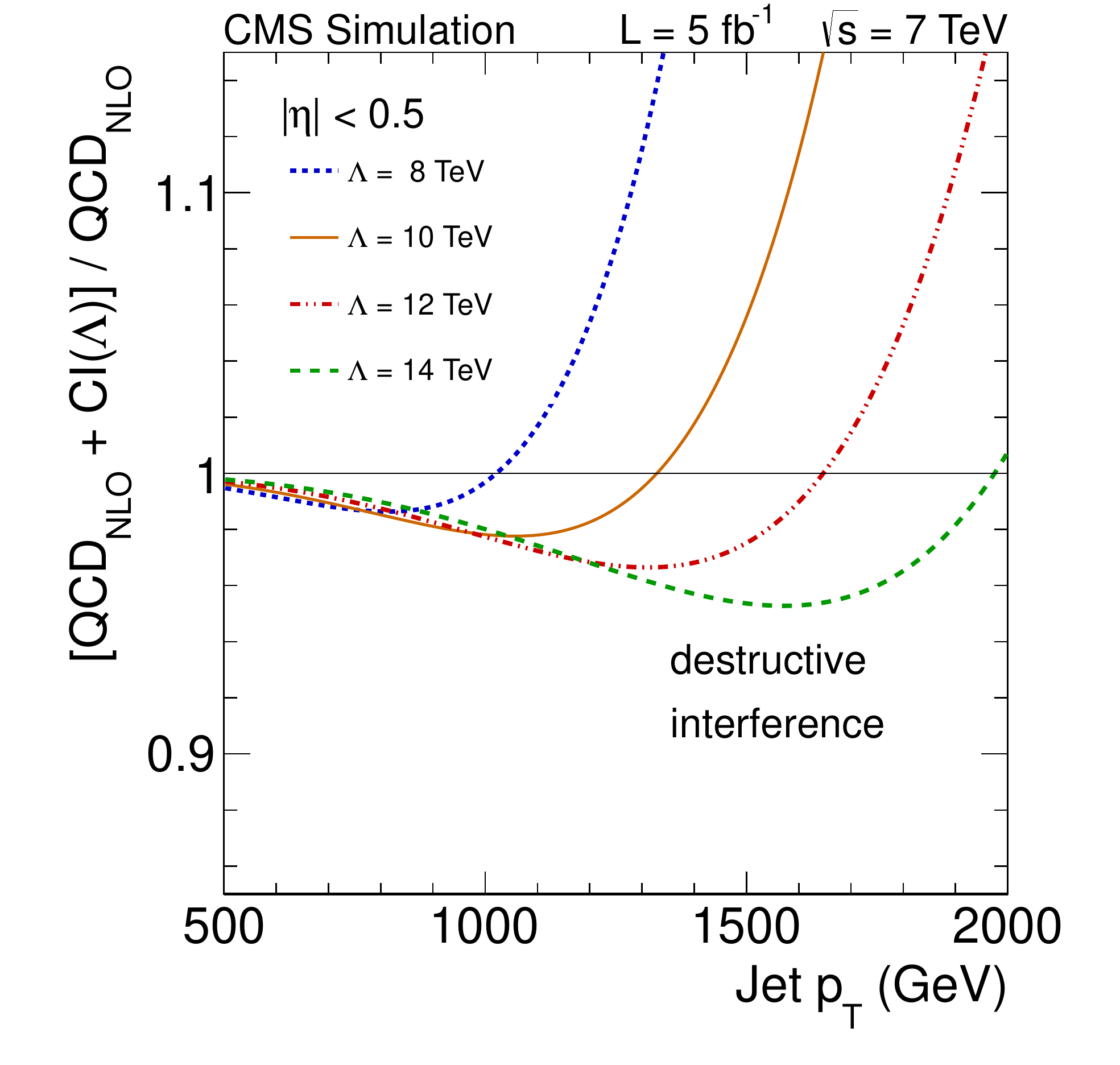}
\includegraphics[width=0.49\textwidth]{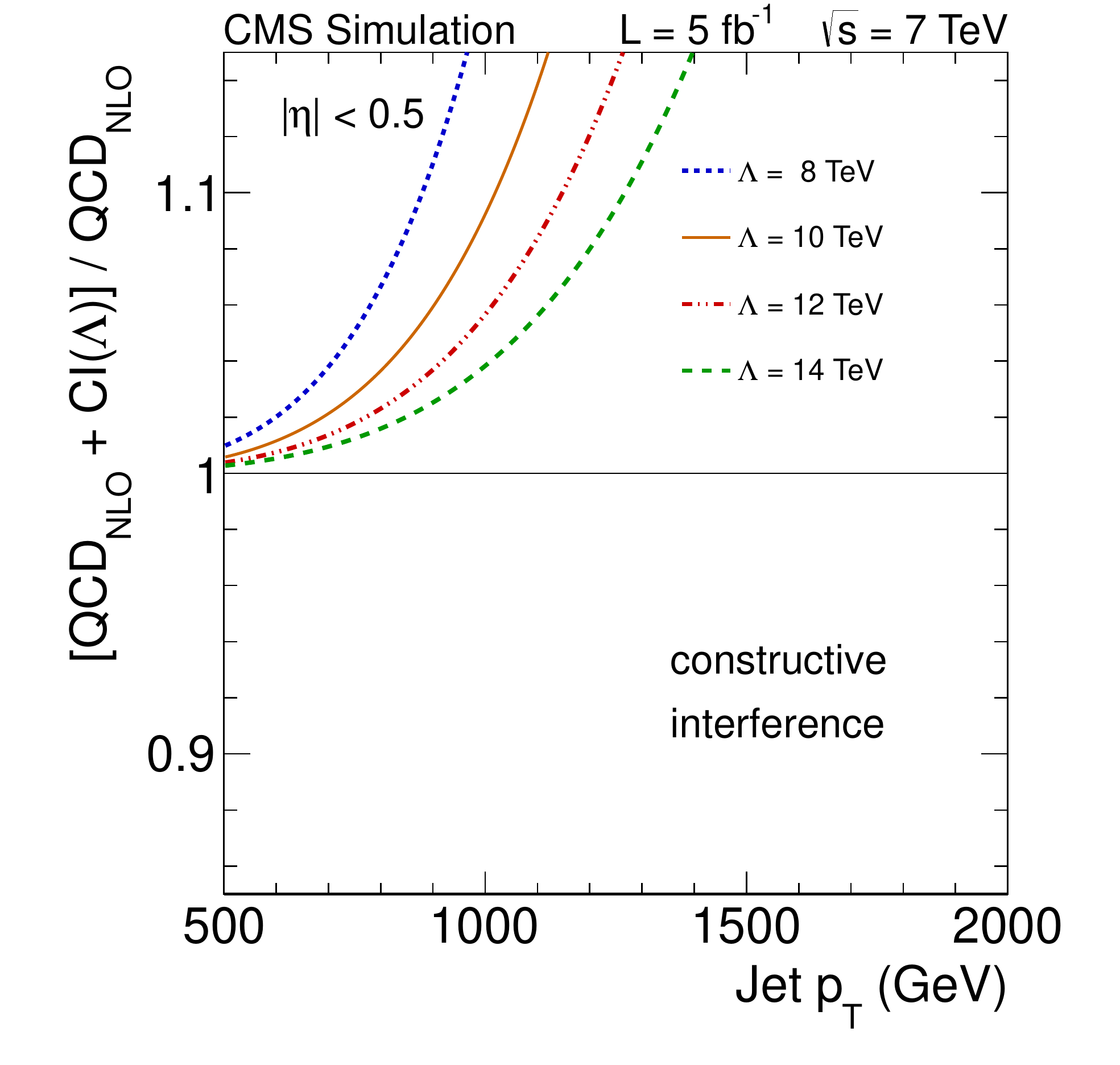}
\caption{
The cross section ratios, $f = [\text{QCD}_\text{NLO} + \text{CI}(\Lambda)]/\text{QCD}_\text{NLO}$, with $\Lambda = 8, 10, 12,$ and 14\TeV,
for models with destructive (\cmsLeft) and constructive (\cmsRight) interference.}
\label{fig:modelratio}
\end{center}
\end{figure}

\begin{figure}[htb!]
\begin{center}
\includegraphics[width=0.49\textwidth]{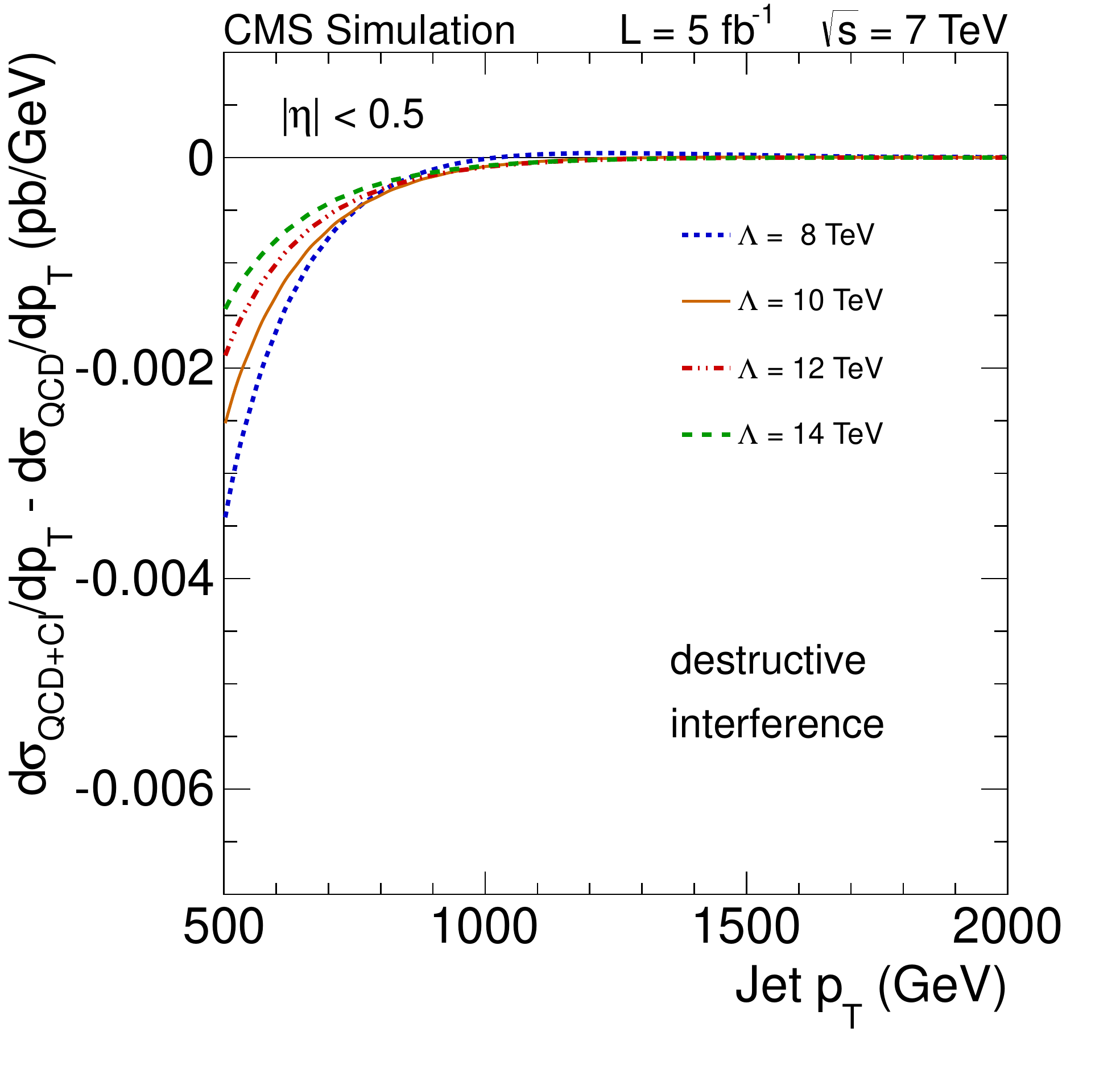}
\includegraphics[width=0.49\textwidth]{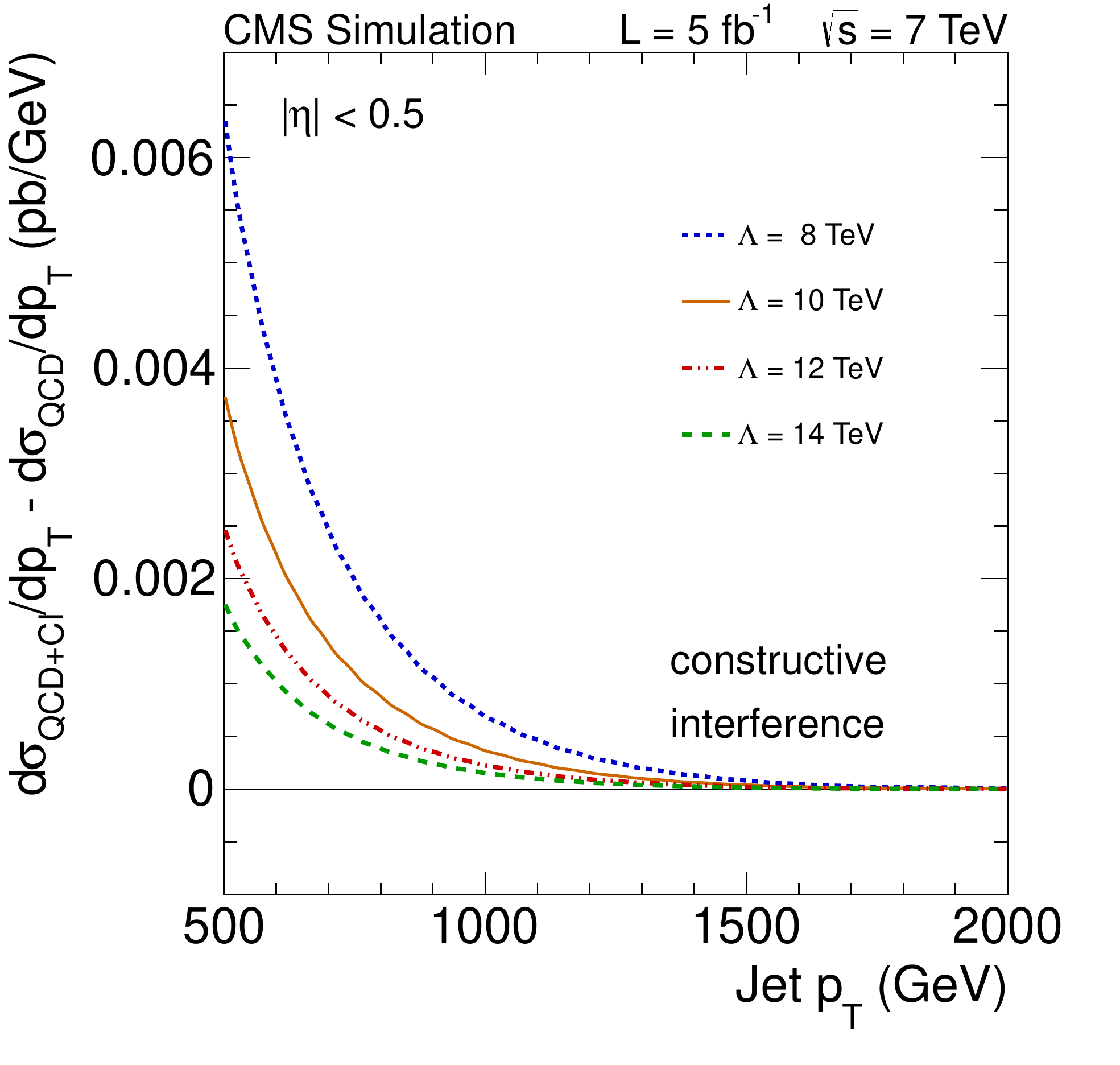}
\caption{
The CI signal spectra, defined as $d\sigma_\text{QCD+CI}/d\pt - d\sigma_\text{QCD}/d\pt$~(pb/GeV)
with $\Lambda = 8, 10, 12$ and 14\TeV, for models
with destructive (\cmsLeft) and constructive (\cmsRight) interference.}
\label{fig:modelsignal}
\end{center}
\end{figure}

The jet $\pt$ dependence of $\text{CI}(\Lambda)$ is modeled by fitting the
ratio $f = [\text{QCD}_\text{NLO} + \text{CI}(\Lambda)] / \text{QCD}_\text{NLO}$
simultaneously to four \PYTHIA CI models with $\Lambda = 3, 5, 8,$ and 12\TeV.
The fit is performed in this manner in order to construct a smooth
interpolation over the four cross section ratios.  Several functional forms
were investigated that gave satisfactory fits, including the
ansatz~\cite{JOwens}:
\ifthenelse{\boolean{cms@external}}
{
\begin{equation}\begin{split}
f = 1 &+ p_1 \, \left(\frac{\pt}{100\GeV}\right)^{p_2} \,\left( \frac{\lambda}{1\TeVns^{-2}}\right)\\
&+ p_3 \, \left(\frac{\pt}{100\GeV} \right)^{p_4} \, \left(\frac{\lambda}{1\TeVns^{-2}}\right)^2.
\label{eq:f}
\end{split}
\end{equation}
}
{
\begin{equation}
f = 1 + p_1 \, \left(\frac{\pt}{100\GeV}\right)^{p_2} \,\left( \frac{\lambda}{1\TeVns^{-2}}\right)+ p_3 \, \left(\frac{\pt}{100\GeV} \right)^{p_4} \, \left(\frac{\lambda}{1\TeVns^{-2}}\right)^2.
\label{eq:f}
\end{equation}
}
In a generator-level study, we verified the adequacy of the extrapolation of Eq.~(\ref{eq:f}) up to 25\TeV.
The results of fitting Eq.~(\ref{eq:f}) to models with
destructive interference are shown in Figure~\ref{fig:fit}.  The fit shown in Fig.~\ref{fig:fit} uses
the central values of the JES, JER, and PDF parameters and
the renormalization ($\mu_\text{r}$) and factorization ($\mu_\text{f}$) scales
set to $\mu_\text{r} = \mu_\text{f} = \text{jet} \, \pt$. Models with constructive interference are
obtained by reversing the sign of the parameter $p_1$.
The fit parameters are given in Table~\ref{tab:fit}.
\begin{table}[htb!]
        \centering
        \topcaption{The fit parameters associated with Fig.~\ref{fig:fit}. The first row lists the
        values of the parameters $p_1, p_2, p_3,$ and $p_4$, while the remaining rows
        list the elements of the associated
        covariance matrix.}
          \begin{scotch}{rrrrr}
            \multicolumn{1}{c}{}         &
             \multicolumn{1}{c}{$p_1$}      &
             \multicolumn{1}{c}{$p_2$}     &
             \multicolumn{1}{c}{$p_3$}    &
             \multicolumn{1}{c}{$p_4$}\\
            \hline
                  & $-1.5 \times 10^{-3}$ & \multicolumn{1}{c}{$3.6$} & $ 1.9 \times 10^{-3}$ & \multicolumn{1}{c}{$ 5.2$}\\
$p_1$      & $ 1.4 \times 10^{-6}$                &$ 3.6 \times 10^{-4}$
       & $-3.4 \times 10^{-7}$          &$ 6.8 \times 10^{-5}$\\
$p_2$    & $ 3.6 \times 10^{-4}$                &$ 9.2 \times 10^{-2}$
       & $-8.4 \times 10^{-5}$                &$ 1.7 \times 10^{-2}$\\
$p_3$    & $-3.4 \times 10^{-7}$                &$-8.4 \times 10^{-5}$
       & $1.0\times 10^{-7}$                &$-2.0 \times 10^{-5}$\\
$p_4$     & $ 6.8 \times 10^{-5}$                &$ 1.7 \times 10^{-2}$
       & $-2.0 \times 10^{-5}$                &$ 4.1 \times 10^{-3}$\\
          \end{scotch}
        \label{tab:fit}
\end{table}
Figures~\ref{fig:modelratio} and \ref{fig:modelsignal} show model
spectra in the jet $\pt$ range
$500 \leq \pt \leq 2000$\GeV for values of $\Lambda$ that are close to the limits  reported in this paper. Figure~\ref{fig:modelratio}
shows that the jet production cross section is enhanced at
sufficiently high jet $\pt$.  However, for interactions that interfere
destructively, the cross section can decrease relative
to the NLO QCD prediction. For example, for $\Lambda = 10$\TeV, the
$\text{QCD}_\text{NLO}$+$\text{CI}$ cross section is lower
than the $\text{QCD}_\text{NLO}$ cross section for jet $\pt < 1.3$\TeV.
Figure~\ref{fig:modelsignal} shows the contact interaction
signal, $\text{CI}(\Lambda)$, as a function of jet $\pt$.

\section{Experimental setup}
\label{sec:exp}
The CMS coordinate system is right-handed with the origin at the
center of the detector, the $x$ axis directed toward the center of the
LHC ring, the $y$ axis directed upward, and the $z$ axis directed along
the counterclockwise proton beam.  We define
$\phi$ to be the azimuthal angle, $\theta$ to be the polar angle, and
the pseudorapidity to be $\eta \equiv -\ln[\tan(\theta/2)]$.
The central feature of the CMS apparatus is a superconducting
solenoid of 6\unit{m} internal diameter, operating with a magnetic field
strength of 3.8\unit{T}.  Within the field volume are the silicon pixel and
strip trackers and the barrel and endcap calorimeters with $|\eta|<3$. Outside
the field volume, in the forward region, there is an iron/quartz-fiber
hadron calorimeter ($3<|\eta|<5$).
Further details about the CMS detector may be found elsewhere~\cite{PTDR2}.

Jets are built from the five types of reconstructed particles: photons,
neutral hadrons, charged hadrons, muons, and electrons, using the CMS particle-flow
reconstruction method~\cite{JME-10-003-PAS} and the anti-\kt
algorithm with a distance parameter of 0.7~\cite{Blazey2000,
Ellis2008484, Buttar:1092852}.
The jet energy scale correction is derived as a
function of the jet $\pt$ and $\eta$, using a $\pt$-balancing
technique~\cite{JME-10-011-PAS}, and
applied to all components of the jet four momentum.

The results reported are based on data collected using un-prescaled single-jet
triggers with $\pt$ thresholds that were changed in steps from 240
to 300\GeV during the data-taking period.
The trigger thresholds were changed in response
to the increase in instantaneous luminosity.
The jet trigger efficiency is constant, ${\sim}98.8\%$, above ${\sim}400$\GeV, well below
the search region.
Events with hadron calorimeter noise are removed~\cite{CMS-CFT-09-019}
and each selected event must have
a primary vertex within 24\unit{cm}
of the geometric center of the detector along the $z$ axis
and within 0.2\unit{cm} in the transverse $x$-$y$ plane, defined by
criteria described in~\cite{PhysRevLett.107.132001}.
The search is restricted
to $|\eta| < 0.5$ where the effects of contact interactions are
predicted to be the largest~\cite{PhysRevLett.50.811, RevModPhys.56.579, Chiappetta1991489, Lane:1996gr}.
The jet $\pt$
spectrum is divided into 20 $\pt$ bins
in the search region $507 \leq \pt \leq 2116$\GeV, where
the bin width is approximately equal to the jet
resolution $\sigma_{\pt}$ given in Eq.~(\ref{eq:jer}).
No jets are observed above 2000\GeV transverse energy.

\section{Results}
\label{sec:results} 
\begin{figure}[htb!]
\begin{center}
\includegraphics[width=0.49\textwidth]{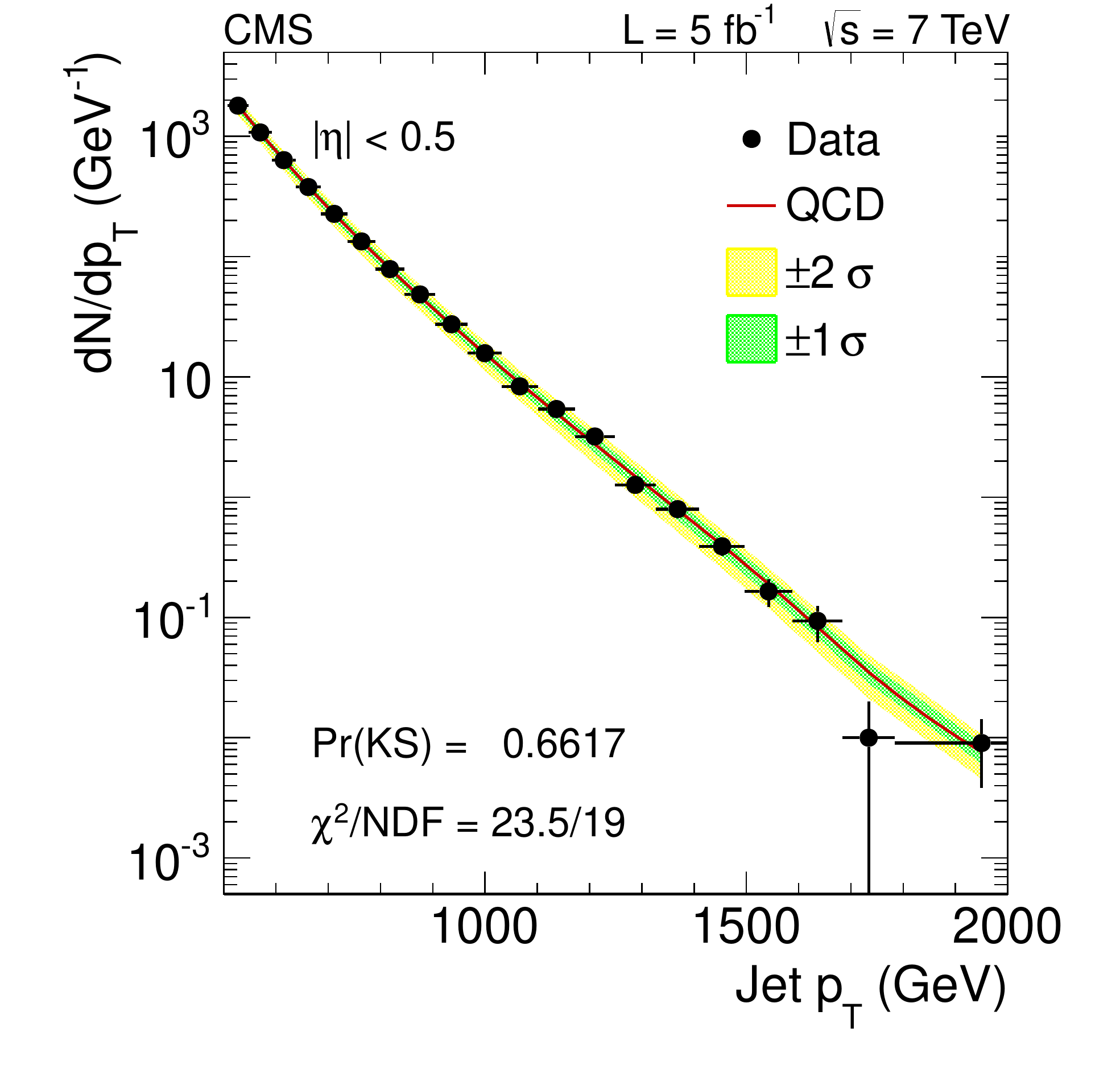}
\includegraphics[width=0.49\textwidth]{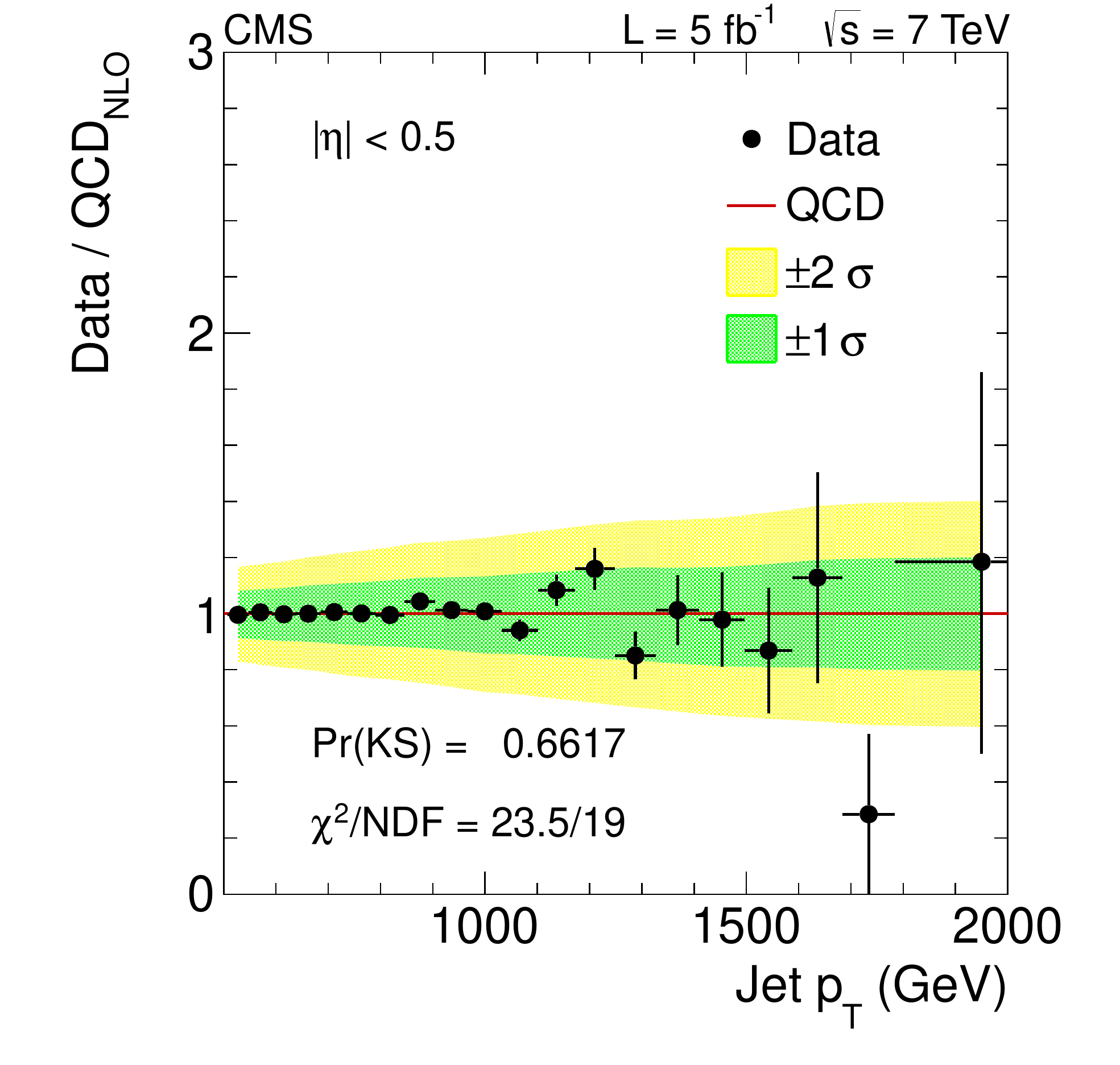}
\caption{The observed jet $\pt$ spectrum
compared with the NLO QCD jet $\pt$ spectrum (\cmsLeft).  The bands represent the total
uncertainty in the prediction and incorporate the uncertainties in
the PDFs, jet energy scale, jet energy
resolution, the renormalization and factorization scales, and the
modeling of the jet $\pt$ dependence of the parameters in
Eq.~(\ref{eq:f}).  The ratio of the
observed to the predicted spectrum (\cmsRight). The error bars represent
the statistical uncertainties in the expected bin count.}
\label{fig:allsyst}
\end{center}
\end{figure}

\begin{figure}[htb!]
\begin{center}
\includegraphics[width=0.49\textwidth]{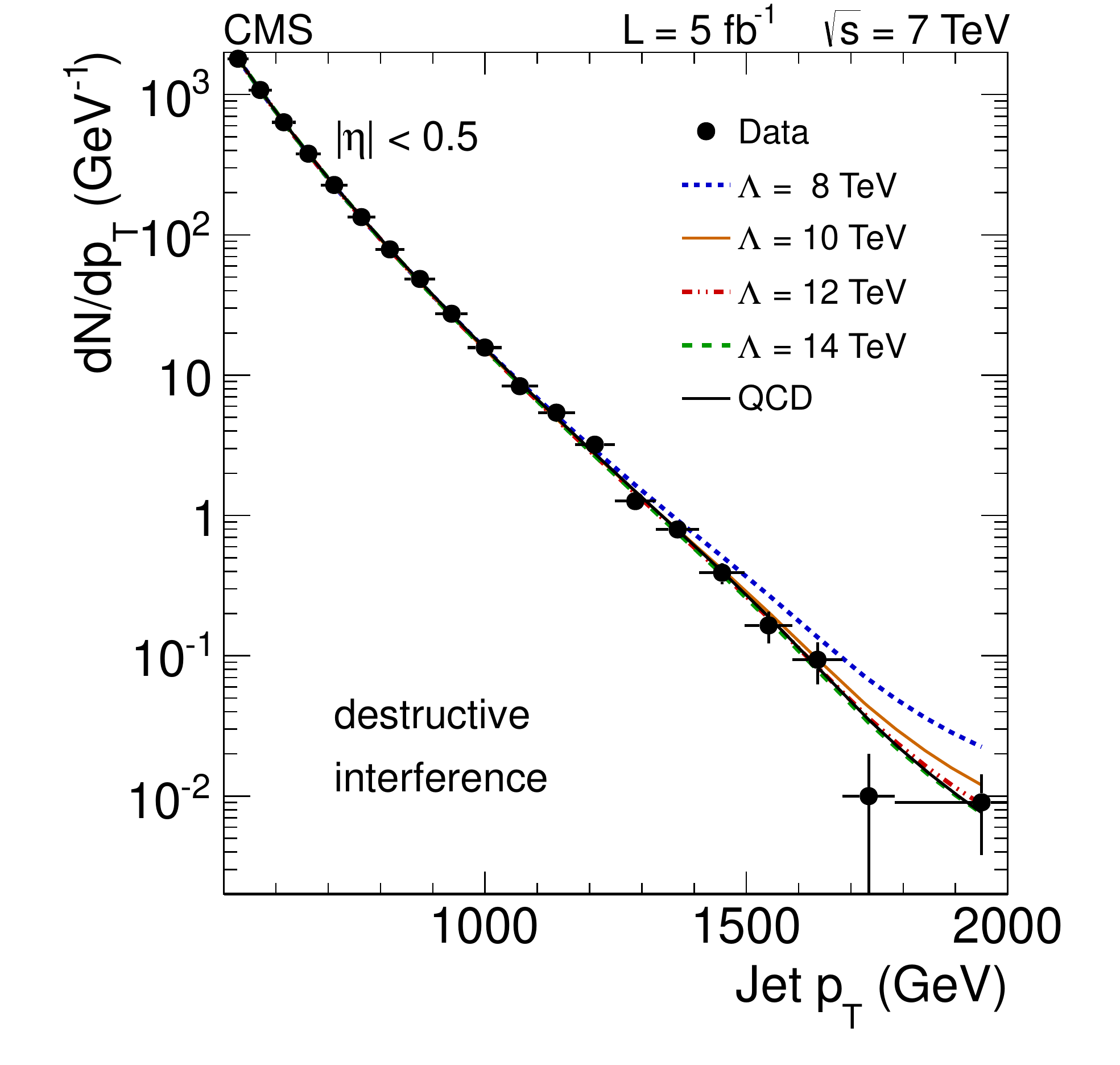}
\includegraphics[width=0.49\textwidth]{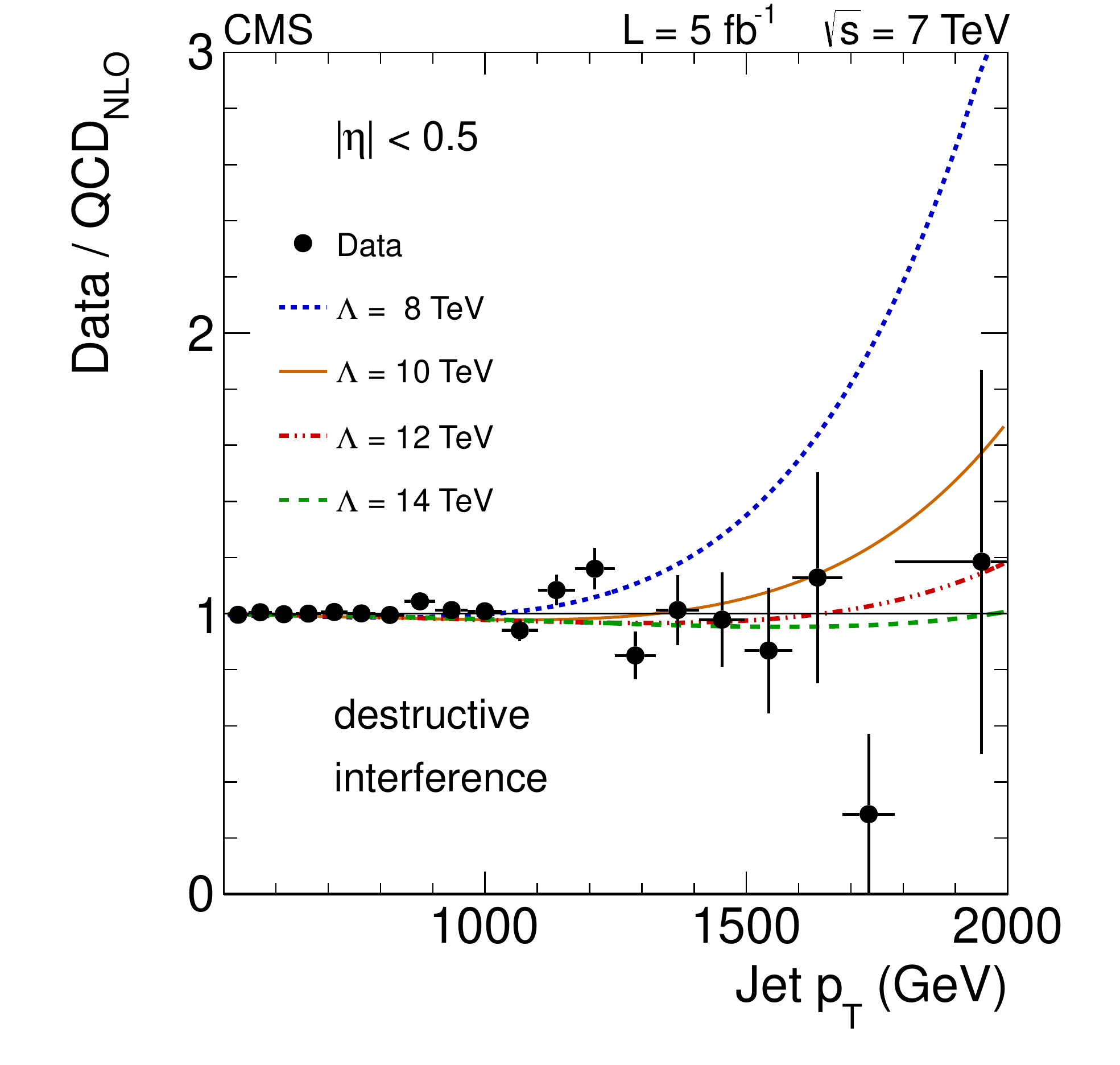}
\caption{The data compared with
model spectra for different values of $\Lambda$ for models with
destructive interference (\cmsLeft).  The
ratio of these spectra to the NLO QCD jet $\pt$ spectrum (\cmsRight).
\label{figure:spectrum_d}
}
\end{center}
\end{figure}

\begin{figure}[htb!]
\begin{center}
\includegraphics[width=0.49\textwidth]{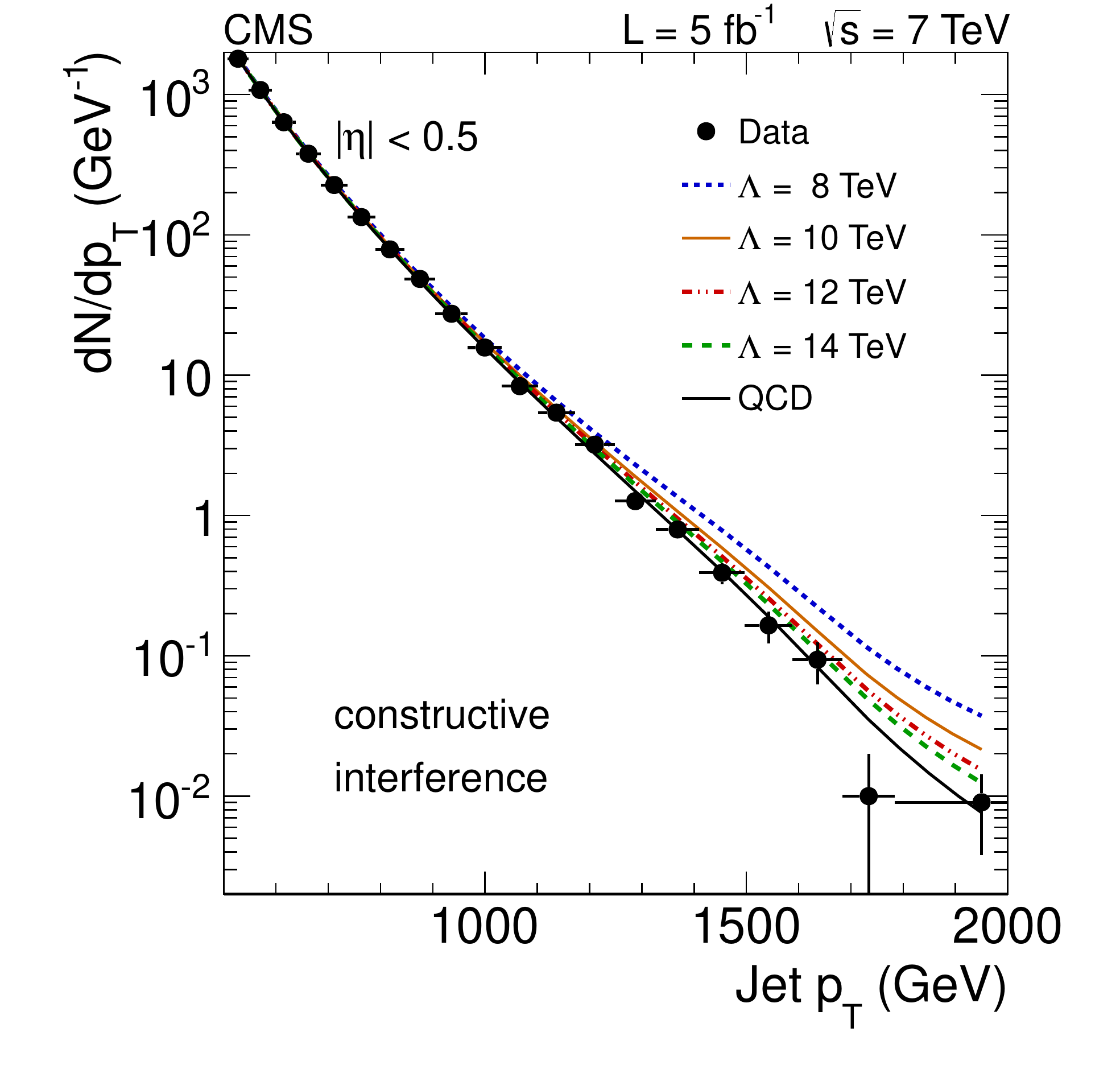}
\includegraphics[width=0.49\textwidth]{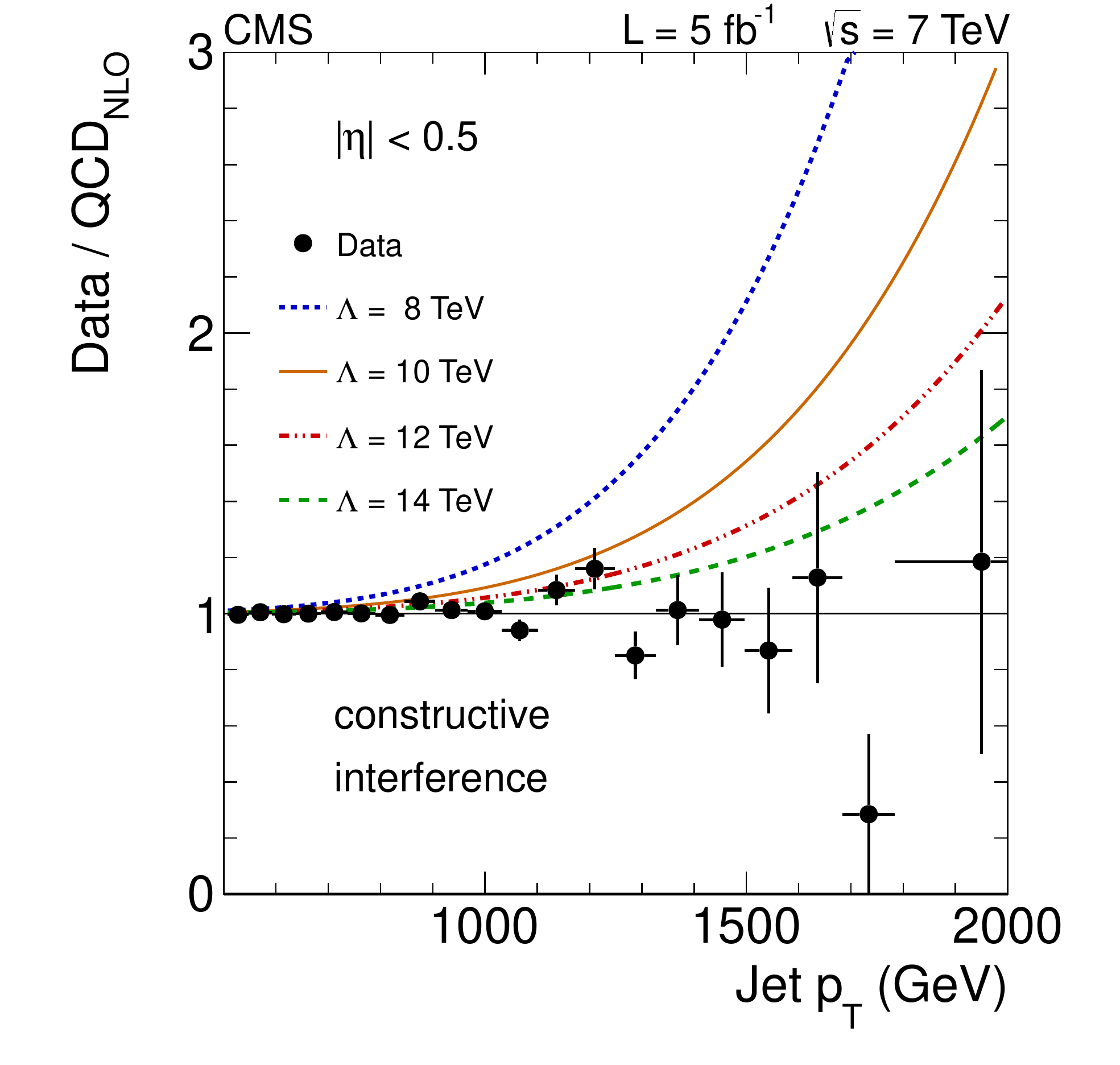}
          \caption{The data compared with model spectra for different
          values of $\Lambda$ for models with constructive interference (\cmsLeft).  The ratio of these
            spectra to the NLO QCD jet $\pt$ spectrum (\cmsRight).}
\label{figure:spectrum_c}
\end{center}
\end{figure}

In Figure~\ref{fig:allsyst} we compare the observed inclusive jet
$\pt$ spectrum with the NLO QCD jet $\pt$ spectrum,
which  is normalized to the total observed jet count in the search region using the normalization
factor $4.007 \pm 0.009\stat\fbinv$ (Section~\ref{sec:int}). The normalization is the ratio of the observed jet count
to the predicted cross section in the search region.
The data and the prediction are in good agreement as indicated by
two standard criteria, the Kolmogorov-Smirnov probability Pr(KS) of 0.66
and the $\chi^2$ per number of degrees of freedom (NDF) of 23.5/19. Table~\ref{tab:eventcounts} lists
the observed jet counts. Figure~\ref{figure:spectrum_d}
compares the observed jet $\pt$ spectrum in the search region with
model spectra for different values of $\Lambda$, for models with
destructive interference.  Figure~\ref{figure:spectrum_c} compares the
data with models with constructive interference.

\begin{table}[htb!]
        \centering
        \topcaption{The observed jet count for each jet $\pt$ bin in the
range 507--2116\GeV.}
  \begin{scotch}{ r  l  r  r l  r }
\multicolumn{1}{c}{Bin} & \multicolumn{1}{c}{$\pt$ (\GeVns{})}  & \multicolumn{1}{c}{Jets} & \multicolumn{1}{c}{Bin} & \multicolumn{1}{c}{$\pt$ (\GeVns{})}  & \multicolumn{1}{c}{Jets} \\
\hline
1 &  507--548 &   73792 & 11 & 1032--1101 &  576  \\
2 &  548--592 &   47416 & 12 & 1101--1172 & 384   \\
3 &  592--638 &   29185 &  13 & 1172--1248 &  243 \\
4 &  638--686 &   18187 &  14 & 1248--1327 & 100  \\
5 &  686--737 &   11565 &  15 & 1327--1410 & 66   \\
6  & 737--790 &  7095  & 16 &1410--1497 & 34 \\
7  & 790--846 &  4413  & 17 &1497--1588 & 15 \\
8   & 846--905 & 2862   & 18 & 1588--1684 & 9 \\
9  &  905--967 &  1699  & 19 & 1684--1784 & 1\\
10 &  967--1032 & 1023  &  20 & 1784--2116 & 3\\
        \end{scotch}
        \label{tab:eventcounts}
\end{table}

\section{Statistical analysis}
\label{sec:int} Since there are no significant deviations between the
observed and predicted spectra, the results are interpreted in terms of
lower limits on the CI scale $\Lambda$ using the models described
in Section~\ref{sec:ci}.  The dominant sources of
systematic uncertainties are associated with the JES, the PDFs,
the JER,
the renormalization ($\mu_\text{r}$) and factorization ($\mu_\text{f}$) scales,
and the modeling parameters of
Eq.~(\ref{eq:f}). Non-perturbative corrections are less than
1\% for transverse momenta above ${\sim}400$\GeV~\cite{PhysRevLett.107.132001}, are negligible compared with
other uncertainties, and are therefore
not applied to our analysis.

In the search region, the inclusive jet spectrum has a range of five orders of
magnitude, which
causes the limits on $\Lambda$ to be sensitive
to the choice of the normalization factor and the size of the data sets.
We have found that a few percent change
in the normalization factor can cause limits to change by as much as 50\%.
Therefore, for the purpose of
computing limits, we have chosen to sidestep the issue of normalization by considering
only the shape of the jet $\pt$ spectrum. This we achieve
by using a multinomial distribution, which is
the probability to observe $K$ counts, $N_j$, $j = 1, \cdots, K$, given the observation
of a total count $N = \sum_{j=1}^K N_j$. The likelihood is then defined by
\begin{eqnarray}
p(D|\lambda, \omega)
  & = & \frac{N!}{N_1! \cdots N_K!} \prod_{j=1}^K \left(\frac{\sigma_j}{\sigma}\right)^{N_j},
  \label{eq:like}
\end{eqnarray}
where $K = 20$ is the number of bins in the search region, $N_j$ is the jet count
in the $j$th jet $\pt$ bin, $D \equiv N_1,\cdots,N_K$,
$\sigma = \sum_{j=1}^K \sigma_j$ and $N$ are the total cross section and total observed count, respectively,
in the search region, and the symbol $\omega$ denotes the
nuisance parameters $p_1,\cdots,p_4$ in Eq.~(\ref{eq:f}).

\begin{figure}[htb!]
\begin{center}
\includegraphics[width=0.49\textwidth]{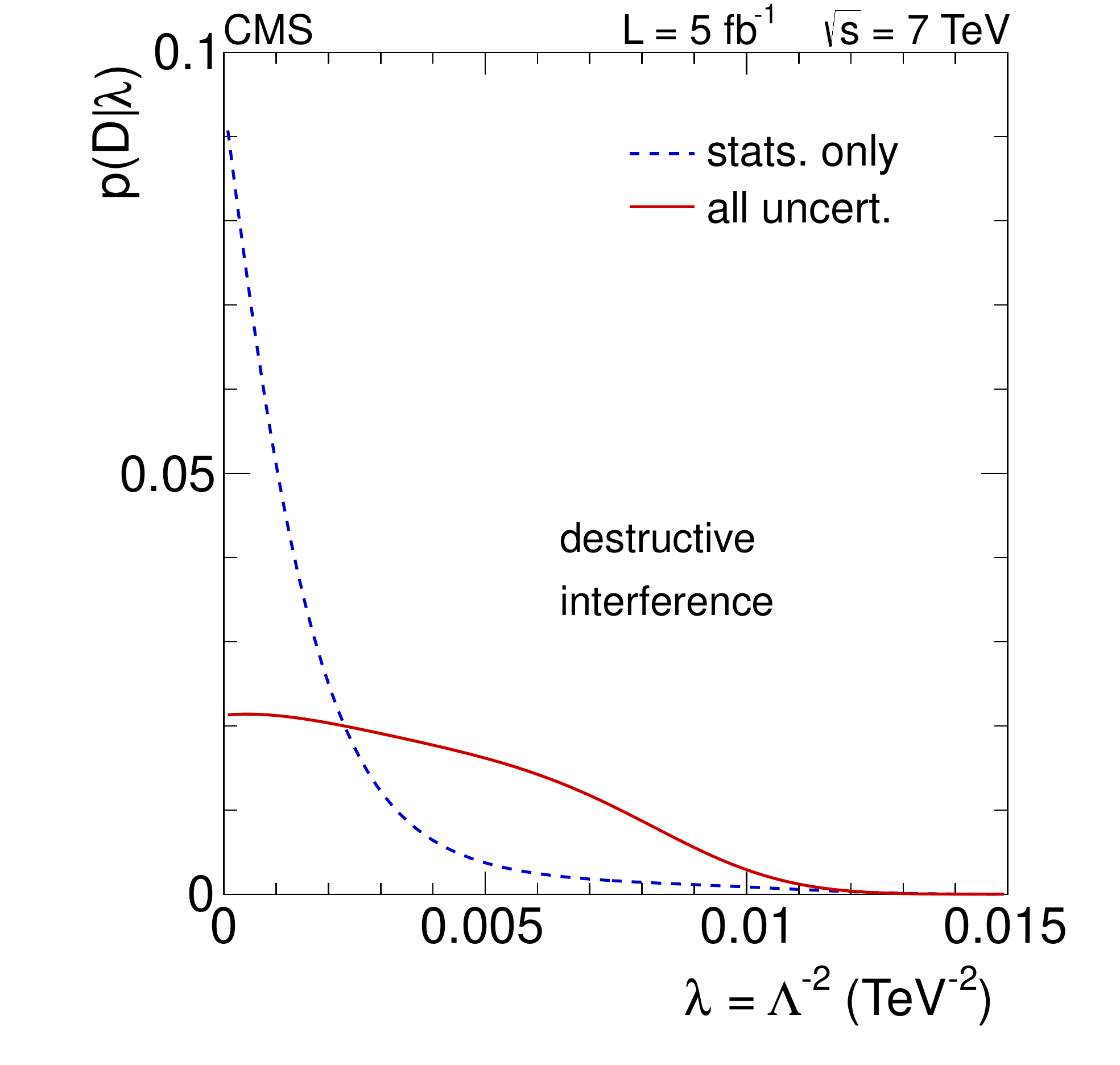}
\includegraphics[width=0.49\textwidth]{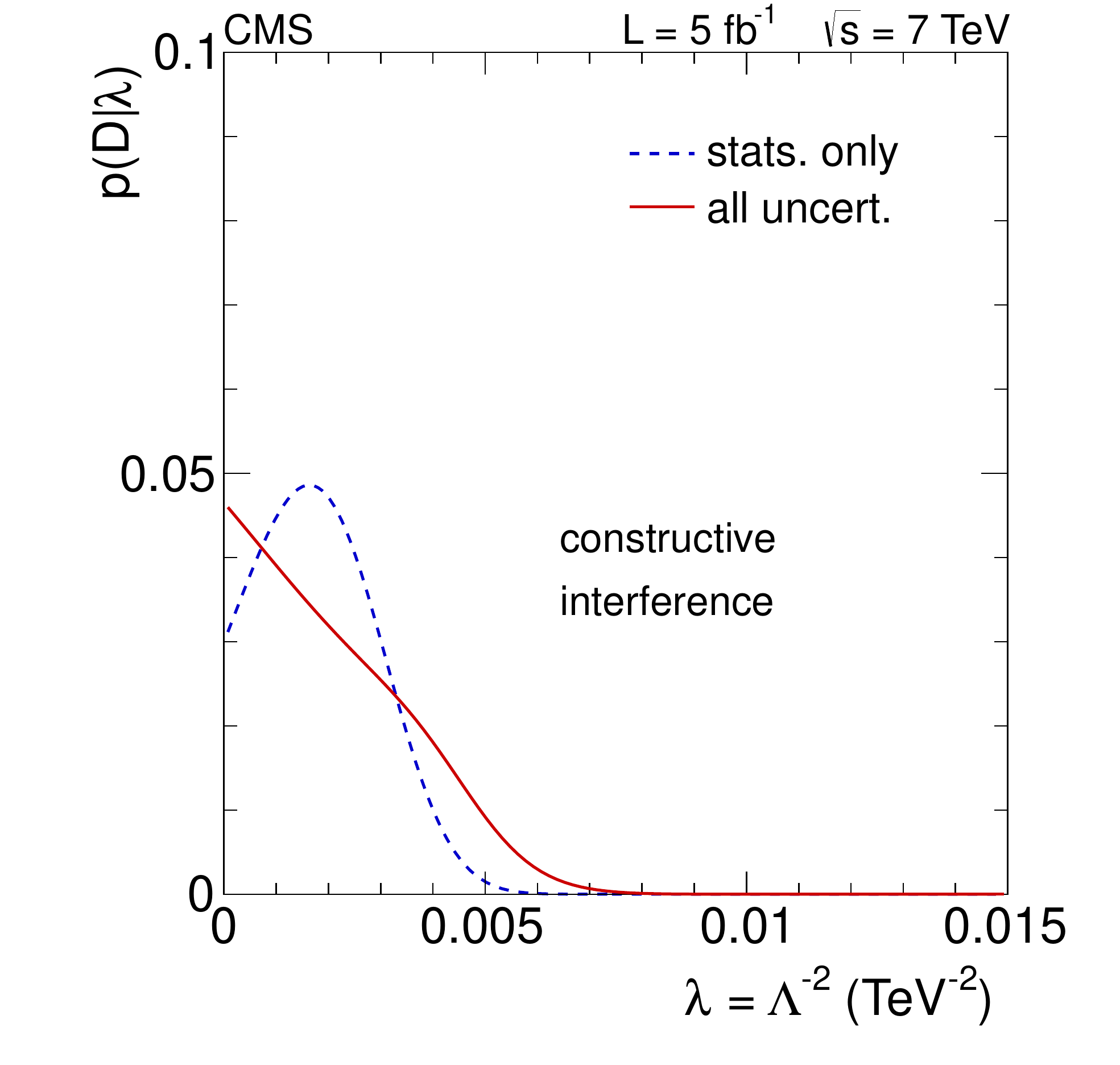}
\caption{The likelihood functions assuming a model with either destructive
                (\cmsLeft) or constructive (\cmsRight) interference.
                The dashed curve is the likelihood function including statistical uncertainties
                only and the central values of all nuisance parameters. The solid curve is the
                likelihood marginalized over all systematic uncertainties.}
\label{fig:likelihoods}
\end{center}
\end{figure}

We account for systematic uncertainties by integrating the likelihood
with respect to a nuisance prior $\pi(\omega)$. In practice,
the likelihood is averaged over the
nuisance parameters, $\omega$, using a discrete
representation of the prior $\pi(\omega)$ constructed
as described in Section~\ref{sec:unc}. This calculation yields the
marginal likelihood $p(D|\lambda) \approx \frac{1}{M} \sum_{m=1}^M p(D|\lambda, \omega_m)$,
where $M$ is the number
of points sampled from the nuisance prior
$\pi(\omega)$ described in Appendix~\ref{sec:prior}, which is the basis of the
limit calculations.
The likelihood functions for models
with destructive and
constructive interference are shown in Figure~\ref{fig:likelihoods}.

\subsection{Uncertainties}
\label{sec:unc}
In principle, a discrete representation of the nuisance prior
$\pi(\omega)$ can be constructed by sampling simultaneously the JES,
JER, PDFs, and the three values of
$\mu_\text{f}$ and $\mu_\text{r}$: $\pt/2$, $\pt$, and $2 \pt$.
However, the CTEQ collaboration~\cite{CTEQ} does not provide a sampling of PDFs.
Instead, CTEQ6.6 contains 44 PDF sets in which the 22 PDF parameters are shifted by approximately
$\pm 1.64$ standard deviations. If we assume the Gaussian approximation
to be valid, we can construct approximate $20 \times 20$
covariance matrices for the jet spectra from the 44 PDF sets. Using these matrices, we generate
ensembles of six correlated spectra: $\text{QCD}_\text{NLO}$,
$\text{QCD}_\text{LO}$, and $(\text{QCD} + \text{CI}(\Lambda))_\text{LO}$ with $\Lambda = 3, 5, 8$, and 12\TeV.
The generation is performed for models both with destructive and constructive interference.
The details of our procedure, which also includes simultaneous sampling of the JES and JER
parameters, are given in Appendix~\ref{sec:prior}.

For a given set of values for the JES, JER, PDF, $\mu_\text{r}$, and $\mu_\text{f}$ parameters,
Eq.~(\ref{eq:f}) is fitted to the
ratio $(\text{QCD}_\text{NLO} + \text{CI})/\text{QCD}_\text{NLO}$ simultaneously
to the four models with $\Lambda = 3, 5, 8$, and 12\TeV. We then sample
a single set of the four nuisance
parameters $\omega = p_1, p_2, p_3, p_4$ from a multivariate Gaussian
using the fitted values
and the associated $4 \times 4$ covariance matrix.
The sampling and
fitting procedure is repeated 500 times, thereby generating a discrete
representation of the nuisance prior $\pi(\omega)$ that incorporates
all uncertainties. We have verified
that our conclusions are robust with respect to variations in the size
of the sample that represents $\pi(\omega)$.

\subsection{Lower limits on \texorpdfstring{$\Lambda$}{Lambda}}
We use the $\text{CL}_\text{s}$ criterion~\cite{CLS, Junk:hep-ex9902006} to compute
upper limits on $\lambda$.  For completeness, we give the details
of these calculations in Appendix~\ref{sec:CLs}. Using the procedure
described in the Appendix, we obtain 95\% lower limits on
$\Lambda$ of 9.9\TeV and 14.3\TeV for models with destructive and
constructive interference, respectively.
These more stringent limits supersede those published by CMS based on a measurement
of the dijet angular distribution~\cite{Chatrchyan:2012bf}. The current search is more sensitive than the
earlier dijet search as evidenced by the expected limits, which for this analysis
are $9.5 \pm 0.6$\TeV and $13.6 \pm 1.6$\TeV, respectively, obtained using 5\fbinv of data.

Limits are also computed with a Bayesian method (Appendix~\ref{sec:Bayesian}) using the marginal likelihood
$p(D|\lambda)$ and two different
priors for $\lambda$: a prior flat in $\lambda$ and a reference prior
~\cite{bib:Bernardo, bib:Sun, bib:refprior}.  Using a flat prior, we find lower limits
on $\Lambda$ of 10.6\TeV and 14.6\TeV at 95\% confidence level
for models with destructive and constructive interference,
respectively. The corresponding limits using the reference prior are
10.1\TeV and 14.1\TeV at 95\% confidence level, respectively.

\section{Summary}
\label{sec:conclusion}
The inclusive jet $\pt$ spectrum of 7\TeV proton-proton
collision events in the ranges $507 \leq \pt \leq 2116$\GeV and $|\eta|
< 0.5$ has been studied using a data set corresponding to an
integrated luminosity of 5.0\fbinv.  The observed jet $\pt$
spectrum is found to be in agreement
with the jet $\pt$ spectrum predicted using perturbative
QCD at NLO when the predicted
spectrum is convolved with the CMS jet response function and
normalized to the observed spectrum in the search region. Should additional interactions
exist that can be modeled as contact interactions with either
destructive or constructive interference, their scale $\Lambda$ is
above 9.9\TeV and 14.3\TeV, respectively, at 95\% confidence level.
We plan to extend this study to the full 8\TeV CMS data set, making use of a recently released program~\cite{GaoCI} to calculate at next-to-leading order the inclusive jet $p_\text{T}$ spectrum
with contact interactions.

It is noteworthy that the
limits reported in this paper, which are the most sensitive
limits published to date, have been obtained reprising the classic
method to search for contact interactions: namely, searching for
deviations from QCD at high jet transverse momentum.

\section*{Acknowledgments}
\label{sec:acknowledgment}
We congratulate our colleagues in the CERN accelerator departments for
the excellent performance of the LHC and thank the technical and
administrative staffs at CERN and at other CMS institutes for their
contributions to the success of the CMS effort. In addition, we
gratefully acknowledge the computing centers and personnel of the
Worldwide LHC Computing Grid for delivering so effectively the
computing infrastructure essential to our analyses. Finally, we
acknowledge the enduring support for the construction and operation of
the LHC and the CMS detector provided by the following funding
agencies: BMWF and FWF (Austria); FNRS and FWO (Belgium); CNPq, CAPES,
FAPERJ, and FAPESP (Brazil); MEYS (Bulgaria); CERN; CAS, MoST, and
NSFC (China); COLCIENCIAS (Colombia); MSES (Croatia); RPF (Cyprus);
MoER, SF0690030s09 and ERDF (Estonia); Academy of Finland, MEC, and
HIP (Finland); CEA and CNRS/IN2P3 (France); BMBF, DFG, and HGF
(Germany); GSRT (Greece); OTKA and NKTH (Hungary); DAE and DST
(India); IPM (Iran); SFI (Ireland); INFN (Italy); NRF and WCU (Korea);
LAS (Lithuania); CINVESTAV, CONACYT, SEP, and UASLP-FAI (Mexico); MSI
(New Zealand); PAEC (Pakistan); MSHE and NSC (Poland); FCT (Portugal);
JINR (Armenia, Belarus, Georgia, Ukraine, Uzbekistan); MON, RosAtom,
RAS and RFBR (Russia); MSTD (Serbia); SEIDI and CPAN (Spain); Swiss
Funding Agencies (Switzerland); NSC (Taipei); ThEP, IPST and NECTEC
(Thailand); TUBITAK and TAEK (Turkey); NASU (Ukraine); STFC (United
Kingdom); DOE and NSF (USA).

\bibliography{auto_generated}   

\appendix

\section{\PYTHIA 6.422 contact interaction configuration}
\label{sec:CIgen}
The scale $\Lambda$ is defined by the CI model in \PYTHIA. In order to facilitate the
re-interpretation of the results using a different model, we provide the details
of the \PYTHIA configuration in Table~\ref{tab:pythia} for
$\Lambda = 8$\TeV and final state parton transverse momenta, $\hat{p}_\text{T}$,
in the range $170 \leq \hat{p}_\text{T} \leq 230$\GeV.
\begin{center}
\begin{table}[htp]
\centering
\topcaption{\PYTHIA 6.422 configuration for $\Lambda = 8$\TeV contact interactions.\label{tab:pythia}}
        \begin{scotch}{   l  c  }
          \multicolumn{2}{c}{\PYTHIA 6.422 settings specific to contact interactions}\\
          \hline
          Settings  & Description \\
          \hline
          ITCM(5)=2          & Switch on contact int. for all quarks \\
          RTCM(41)=8000  & Set contact scale $\Lambda$ to 8\TeV \\
          RTCM(42)=1         & Sign of contact int. is + \\
          MSUB(381)=1    & $\cPq_i \cPq_j \rightarrow  \cPq_i \cPq_j$ via QCD plus a contact int. \\
          MSUB(382)=1    & $q_i \cPaq_i \rightarrow q_k \cPaq_k$ via QCD plus a contact int. \\
          MSUB(13)=1         & $\cPq_i \cPaq_i \rightarrow \Pg \Pg$ via normal QCD \\
          MSUB(28)=1         & $\cPq_i \Pg \rightarrow \cPq_i \Pg$  via normal QCD \\
          MSUB(53)=1         & $\Pg \Pg \rightarrow \cPq_k \cPaq_k$ via normal QCD \\
          MSUB(68)=1         & $\Pg \Pg \rightarrow \Pg \Pg$ via normal QCD \\
          CKIN(3)=170        & minimum $\hat{p}_T$ for hard int. \\
          CKIN(4)=230        & maximum $\hat{p}_T$ for hard int. \\
        \end{scotch}
\end{table}
\end{center}

\section{Statistical details}
\subsection{Nuisance prior}
\label{sec:prior}
We approximate the nuisance prior $\pi(\omega)$ starting with
two sets of ensembles. In the first, the six 20-bin model spectra
$\text{QCD}_\text{NLO}$, $\text{QCD}_\text{LO}$,
and $[\text{QCD} + \text{CI}(\Lambda)]_\text{LO}$ with $\Lambda = 3, 5, 8$, and 12\TeV
are varied,
reflecting random variations in the PDF parameters as well as random choices
of the three $\mu_\text{r}$ and $\mu_\text{f}$ scales, while keeping the JES and JER parameters
fixed at their central values;  we call these the PDF ensembles. In the second set of ensembles, the
JES and JER
parameters are varied simultaneously, while keeping the PDF parameters fixed to their central
values and the renormalization and factorization scales at their nominal values;
we call these the JES/JER ensembles.

\subsubsection{Generating the PDF ensembles}
\label{sec:pdfens}
In the PDF ensembles,
each of the six model spectra
is sampled from a multivariate Gaussian distribution
using the associated $20 \times 20$ covariance matrix.
For each model spectrum, the
covariance matrix is approximated by
\begin{equation}
        C_{nm} = \sum_{i=1}^{22} \sum_{j=1}^{22} \Delta X_{ni} \, \Delta X_{mj},
\end{equation}
where $\Delta X_{ni} = (X^+_{ni} - X^-_{ni})/2$ and $X^\pm_{ni}$ are the cross
section values for $n$th jet bin associated with the $+$ and $-$ variations of the $i$th pair
of CTEQ6.6 PDF sets. CTEQ~\cite{CTEQ} publishes approximate 90\%
intervals. We therefore approximate 68\% intervals by dividing each $\Delta X$
by 1.64. The correlation induced by the PDF uncertainties across all six model spectra
is maintained by using the same set of underlying Gaussian variates during the sampling
of the spectra.

\subsubsection{Generating the JES/JER ensembles}
\label{sec:jesjerens}
In the JES/JER ensembles, the JES and JER parameters are sampled simultaneously
for the five model spectra $\text{QCD}_\text{LO}$,
and $(\text{QCD} + \text{CI})_\text{LO}$ with $\Lambda = 3, 5, 8$, and 12\TeV, yielding
ensembles of correlated shifts from the central
JES, JER, and PDF  values of the $\text{QCD}_\text{LO}$ and $(\text{QCD+CI})_\text{LO}$ spectra.
For example, we
compute the spectral residuals $\delta\sigma = \text{QCD}^\prime - \text{QCD}_\text{central}$,
where $\text{QCD}^\prime$ is the shifted jet $\pt$ spectrum and
$\text{QCD}_\text{central}$ is the jet $\pt$ spectrum computed using the central values
of the JES, JER, and PDF parameters.
Coherent shifts of the jet energy scale are calculated for every jet
in every simulated event. The jet $\pt$ is shifted by $x \delta $ for
each component of the jet energy scale uncertainty, of which there are
sixteen, where $x$ is a Gaussian variate of zero mean and unit
variance, and $\delta$ is a jet-dependent uncertainty for a given
component.  The contributions from all uncertainty components are
summed to obtain an overall shift in the jet $\pt$. From studies of
dijet asymmetry and photon+jet $\pt$ balancing, the uncertainty in the
jet energy resolution is estimated to be 10\% in the pseudorapidity range
$|\eta| < 0.5$~\cite{PhysRevLett.107.132001}.
We sample the jet energy resolution using a procedure identical to
that used to sample the jet energy scale, but using a single Gaussian
variate.

\subsubsection{Generating the JES/JER/PDF ensemble}
Another ensemble is created, from the PDF ensembles
and the JES/JER ensembles, that approximates simultaneous
sampling from the JES, JER, PDF, renormalization, and factorization
parameters.  We pick at random
a correlated set of six spectra
from the PDF ensembles, and a correlated set of five spectral residuals  from
the JES/JER ensembles. The JES/JER spectral residuals $\delta\sigma$ are
added to the corresponding shifted spectrum from the
PDF ensembles, thereby creating a spectrum in which the JES, JER,
PDF, $\mu_\text{r}$, and $\mu_\text{f}$ parameters have been randomly shifted.
The NLO QCD spectrum (from the PDF ensembles) is shifted using the LO QCD
JES/JER spectral residuals in order to approximate the effect of the JES and JER
uncertainties in this spectrum.

The result of the above procedure is an ensemble of sets of properly correlated spectra
$\text{QCD}_\text{NLO} + \text{CI}(\Lambda)$ with $\Lambda = 3, 5, 8,$ and 12\TeV, in which the JES, JER,
PDF, $\mu_\text{r}$ and $\mu_\text{f}$ parameters vary randomly. The ansatz in Eq.~(\ref{eq:f})
is then fitted to the quartet of ratios  $[\text{QCD}_\text{NLO} + \text{CI}(\Lambda)]$ $/$ $\text{QCD}_\text{NLO}$
as described in Section~\ref{sec:unc} to
obtain  parameter values for $p_1, p_2, p_3,$ and $p_4$.
Five hundred sets of these parameters are generated, constituting a
discrete approximation to the prior $\pi(\omega) \equiv \pi(p_1, p_2, p_3, p_4)$.

\subsection{\texorpdfstring{CL$_\text{s}$}{CLs} calculation}
\label{sec:CLs}
Since CL$_\text{s}$ is a criterion rather than a method, it is necessary to document exactly how
a CL$_\text{s}$ limit is calculated. Such a
calculation requires two elements: a test statistic $Q$ that
depends on the quantity of interest and its sampling distribution
for two different hypotheses, here $\lambda > 0$, which we denote by $H_\lambda$, and
$\lambda = 0$, which we denote by $H_0$. $H_\lambda$ is the signal plus background
hypothesis while $H_0$ is the background-only hypothesis. For this study, we use the statistic
\begin{equation}
Q(\lambda) = t(D, \lambda) \equiv -2 \ln\left[p(D|\lambda) / p(D|0)\right],
\end{equation}
where $p(D|\lambda)$ is the marginal likelihood
\begin{equation}
\begin{split}
  p(D|\lambda) & =  \int p(D|\lambda, \omega) \, \pi(\omega) \, \rd\omega, \\
          & \approx  \frac{1}{M} \sum_{m=1}^M p(D|\lambda, \omega_m),
          \label{eq:mlike}
\end{split}
\end{equation}
where $M = 500$ is the number of points $\omega = p_1, p_2, p_3, p_4$ sampled from
the nuisance prior $\pi(\omega)$ described in Appendix~\ref{sec:prior}.
We compute the sampling distributions
\begin{align}
p(Q |H_\lambda)         & =  \int \delta[Q - t(D, \lambda)] \,  p(D|\lambda) \, \rd D,\\
\intertext{and}
p(Q|H_0)             & =  \int \delta[Q - t(D, \lambda)] \,  p(D|0) \, \rd D,
\end{align}
pertaining to the hypotheses $H_\lambda$ and $H_0$, respectively, and solve
\begin{equation}
\text{CL}_\text{s} \equiv p(\lambda) / p(0) = 0.05,
\end{equation}
to obtain a 95 \% confidence level (CL$_\text{s}$) upper limit on $\lambda$,
where the p-value $p(\lambda)$ is defined by
\begin{equation}
p(\lambda) = \mbox{Pr}[Q(\lambda) > Q_0(\lambda)],
\end{equation}
and $Q_0$ is the observed value of  $Q$.

In practice, the CL$_\text{s}$ limits are approximated as follows:
\begin{enumerate}
\item Choose a value of $\lambda$, say $\lambda^*$,
and compute the observed value of $Q$, $Q_0(\lambda^*)$.
\item Choose at random one of the $M = 500$ sets of nuisance parameters $p_1, p_2, p_3$, and
$p_4$.
\item Generate a spectrum of $K = 20$ counts, $D$,
according to the multinomial distribution, Eq.~(\ref{eq:like}), with
$\lambda = \lambda^*$, which
corresponds to the hypothesis $H_\lambda$. Compute $Q = t(D, \lambda^*)$ and
keep track of how often $Q(\lambda^*) > Q_0(\lambda^*)$. Call this count $n_\lambda$.
\item Generate another set of 20 counts, $D$, but with $\lambda  = 0$, corresponding to
the hypothesis $H_0$. Compute $Q = t(D, \lambda^*)$ and
keep track of how often $Q(\lambda^*) > Q_0(\lambda^*)$. Call this count $n_0$.
\item Repeat $25,000$ times steps 2 to 4, compute $\text{CL}_\text{s} \approx n_\lambda / n_0$
and report $\lambda = \lambda^*$
as the upper limit on $\lambda$ at 95\% CL if $\text{CL}_\text{s}$ is sufficiently close to 0.05;
otherwise, keep repeating steps 1 to 4 with different values of $\lambda$.
The algorithm starts with two values
of $\lambda$ that are likely to bracket the solution and the solution is found using a
binary search, which typically requires about 10 to 15 iterations.
\end{enumerate}

\subsection{Bayesian calculation}
\label{sec:Bayesian}
The Bayesian limit calculations use the marginal likelihood, Eq.~(\ref{eq:mlike}),
and two different (formal) priors $\pi(\lambda)$: a prior
flat in $\lambda$ and a reference prior~\cite{bib:Bernardo, bib:Sun, bib:refprior},
which we calculate numerically~\cite{bib:refprior}.
An upper limit on $\lambda$, $\lambda^*$, is computed by solving
\begin{equation}
  \int_0^{\lambda^*} \, p(D|\lambda) \, \pi(\lambda) \, \rd\lambda / p(D) = 0.95,
\end{equation}
where $p(D)$ is a normalization constant.  The integrals are performed using numerical
quadrature.
\cleardoublepage \appendix\section{The CMS Collaboration \label{app:collab}}\begin{sloppypar}\hyphenpenalty=5000\widowpenalty=500\clubpenalty=5000\textbf{Yerevan Physics Institute,  Yerevan,  Armenia}\\*[0pt]
S.~Chatrchyan, V.~Khachatryan, A.M.~Sirunyan, A.~Tumasyan
\vskip\cmsinstskip
\textbf{Institut f\"{u}r Hochenergiephysik der OeAW,  Wien,  Austria}\\*[0pt]
W.~Adam, E.~Aguilo, T.~Bergauer, M.~Dragicevic, J.~Er\"{o}, C.~Fabjan\cmsAuthorMark{1}, M.~Friedl, R.~Fr\"{u}hwirth\cmsAuthorMark{1}, V.M.~Ghete, N.~H\"{o}rmann, J.~Hrubec, M.~Jeitler\cmsAuthorMark{1}, W.~Kiesenhofer, V.~Kn\"{u}nz, M.~Krammer\cmsAuthorMark{1}, I.~Kr\"{a}tschmer, D.~Liko, I.~Mikulec, M.~Pernicka$^{\textrm{\dag}}$, D.~Rabady\cmsAuthorMark{2}, B.~Rahbaran, C.~Rohringer, H.~Rohringer, R.~Sch\"{o}fbeck, J.~Strauss, A.~Taurok, W.~Waltenberger, C.-E.~Wulz\cmsAuthorMark{1}
% ...
\vskip\cmsinstskip
\dag:~Deceased\\
1:~~Also at Vienna University of Technology, Vienna, Austria\\
2:~~Also at CERN, European Organization for Nuclear Research, Geneva, Switzerland\\
% ...

\end{sloppypar}
\end{document}